# Mid-infrared surface plasmon polariton chemical sensing on fiber-coupled ITO coated glass

JAVIER MARTÍNEZ,[1] AIRÁN RÓDENAS,[1,*] MAGDALENA AGUILÓ,[1] TONEY FERNANDEZ,[2] JAVIER SOLIS,[3] AND FRANCESC DÍAZ[1]

[1] *Departament de Química Física i Inorgànica, Universitat Rovira i Virgili, Tarragona, Spain*
[2] *Dipartimento di Fisica, Politecnico di Milano, Milano, Italy*
[3] *Laser Processing Group, Instituto de Óptica-CSIC, Madrid, Spain*
**Corresponding author: arodenas@gmail.com*

**A novel fiber-coupled ITO coated glass slide sensor for performing surface plasmon polariton chemical monitoring in the ~3.5 µm mid-IR range is reported. Efficient mid-IR fiber coupling is achieved with 3D laser written waveguides, and the coupling of glass waveguide modes to ITO SPPs is driven by the varying phase matching conditions of different aqueous analytes across the anomalous dispersion range determined by their molecular fingerprints. By means of using both a mid-IR fiber supercontinuum source and a diode laser the excitation of SPPs is demonstrated. The sensor sensitivity is tested by discriminating CH from OH features of ethanol in water solutions, demonstrating an instrumental ethanol LOD of 0.02% in a wide concentration range of at least 0–50%. The efficient optical monitoring of mid-IR SPPs in smart glass could have a broad range of applications in biological and chemical sensing.**

Since the introduction of the quantum cascade laser [1], the rapid development of mid-IR technology has fostered research in novel photonic devices for sensing [2]. In the mid-IR range the strong molecular absorption features associated with vibrational resonances allow for distinction of different chemical species within a substance [3]. Moreover, the sensing performance of mid-IR devices can be enhanced by incorporating field enhancement and confinement techniques that effectively increase the strength of the light-analyte interaction [4]. A widely extended mechanism for sensing in the VIS and NIR ranges is the surface plasmon polariton (SPP) [5, 6]. However, the plasma frequency of noble metals typically used in SPP devices lies on the visible part of the spectrum, making them less appealing plasmonic materials at IR wavelengths [7]. For this reason, and since the properties of metallic thin films cannot be tuned, other materials have been considered in recent years, such as semiconductors, silicides, germanides, transparent conductive oxides, or graphene. On the other hand, particular methods like prism interrogation [6, 8], waveguide gratings and tapers [6, 9], or nanoantenna arrays [10], are typically used to access plasmonic modes as normally done in the VIS and NIR regions, but highly efficient direct fiber-coupling schemes are still lacking in the mid-IR.

In this letter, a simple yet robust concept for sensing at mid-IR wavelengths through the excitation of SPPs is explored. The concept is numerically investigated and also fabricated using 3D laser writing (3DLW) of waveguide circuits and thin film sputtering. The experimental observation of mid-IR SPP excitation is presented together with an evaluation of the sensing potential for a specific analyte. The proposed mid-IR SPP sensor consists of a fiber-coupled waveguide inside an ITO covered silica slide. ITO is chosen for its superior plasmonic properties within the spectral range of interest ~3.5 µm [7]. A sketch of the device is shown in Figure 1. The width of the thin film is considered infinite for modelling purposes. Mid-IR light is coupled into the 3DLW waveguide which then routes the light to the analyte interaction area and a detector monitors the output power of the waveguide. An interferometric scheme could be further applied on a Mach-Zehnder configuration although this is out of the scope of the present letter.

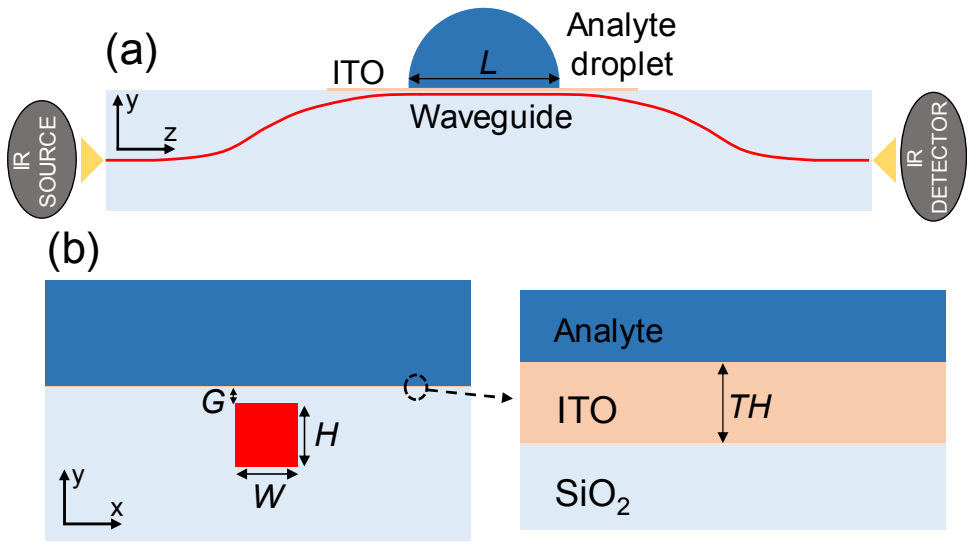

**Fig. 1.** Sketch of the proposed fiber-coupled mid-IR SPP glass sensor (a) and detailed waveguide cross-section design (b).

The mechanism of SPP excitation is the phase matching of the dielectric waveguide mode and the SPP mode as well as the spatial overlapping between the evanescent field of the waveguide mode and the thin film mode [5]. The intense and broad optical absorption of liquid water associated to the stretching of OH groups is accompanied by a large refractive index variation around 3.5 µm, shown in Figure 2, which enables index matching between liquid water and silica glass. This fact ensures the refractive symmetry needed between the *substrate* (silica) and the *superstrate* (water) which, in turn, allows the existence of a SPP mode and hence the aforementioned phase matching. Following the same reasoning, other substances added to the analyte having molecular groups with mid-IR resonances will modify the SPP coupling condition. Note that the presented scheme avoids the need of structuring the surface to achieve phase matching, which notably simplifies the fabrication process.

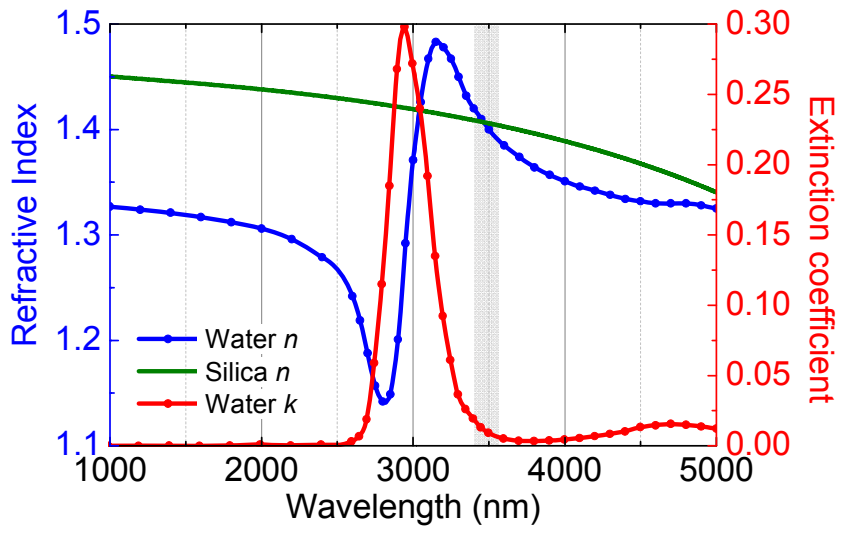

**Fig. 2.** Refractive index of water and fused silica and extinction coefficient of water in the 2 - 4 µm band.

As sketched in Fig. 1, the waveguide parameters are its width and height (*W* and *H*) and its index increase with respect to the unmodified bulk surrounding silica ($\Delta n$). Details of the fabrication and characterization of these 3DLW mid-IR waveguides were recently published [11]. Between the core and the thin film there is also a gap (*G*) of unmodified silica. The interaction length of the generated plasma wave and the liquid is defined as *L* which will be ~2 mm. Regarding the thin film, a thick layer of ITO (*TH*) was deposited onto the silica substrate by DC magnetron sputtering of a commercial ITO target [12]. Optical constants as well as the *TH* of the deposited ITO thin film were retrieved through the fitting of its transmittance data to a hybrid Forouhi-Bloomer plus Drude model [13]. For the particular thin film deposited, these are carrier concentration $N = 13.4\text{x}10^{20}$ cm$^{-3}$, mobility $\mu = 23.4$ cm$^2$/(V·s) and $TH = 54$ nm. From the said model one can extrapolate the complex ITO mid-IR relative permittivity.

With the aid of COMSOL Multiphysics® software the electro-magnetic transversal bound modes of the mid-IR SPP structure were solved via the finite element method (FEM). The mid-IR relative permittivity of the rest of different materials involved (liquid water and silica glass) was taken from literature [14, 15]. For example, at $\lambda = 3.47$ µm they become, respectively, $\varepsilon_{water} = 1.98 + i0.03$, $\varepsilon_{silica} = 1.98$ and $\varepsilon_{ITO} = -28.23 + i11.17$. Note that silica losses are here considered negligible in comparison with water and ITO. Calculated TM mode profiles at $\lambda = 3.49$ µm are shown in Figure 3(a) with the corresponding complex effective index (n + ik) for two different silica gaps and when air or water are covering the chip. It can be seen that a coupled waveguide-SPP mode appears when the superstrate is water, and not air. This coupling becomes the more evident the smaller the gap (*G*), leading to increased mode losses as well. The case $G = 3$ µm is also displayed here since it is a closer value to the one here experimentally reported. The calculated field penetration into the liquid was numerically computed to be 1.6 µm, taking the intensity as an exponentially decaying function [6].

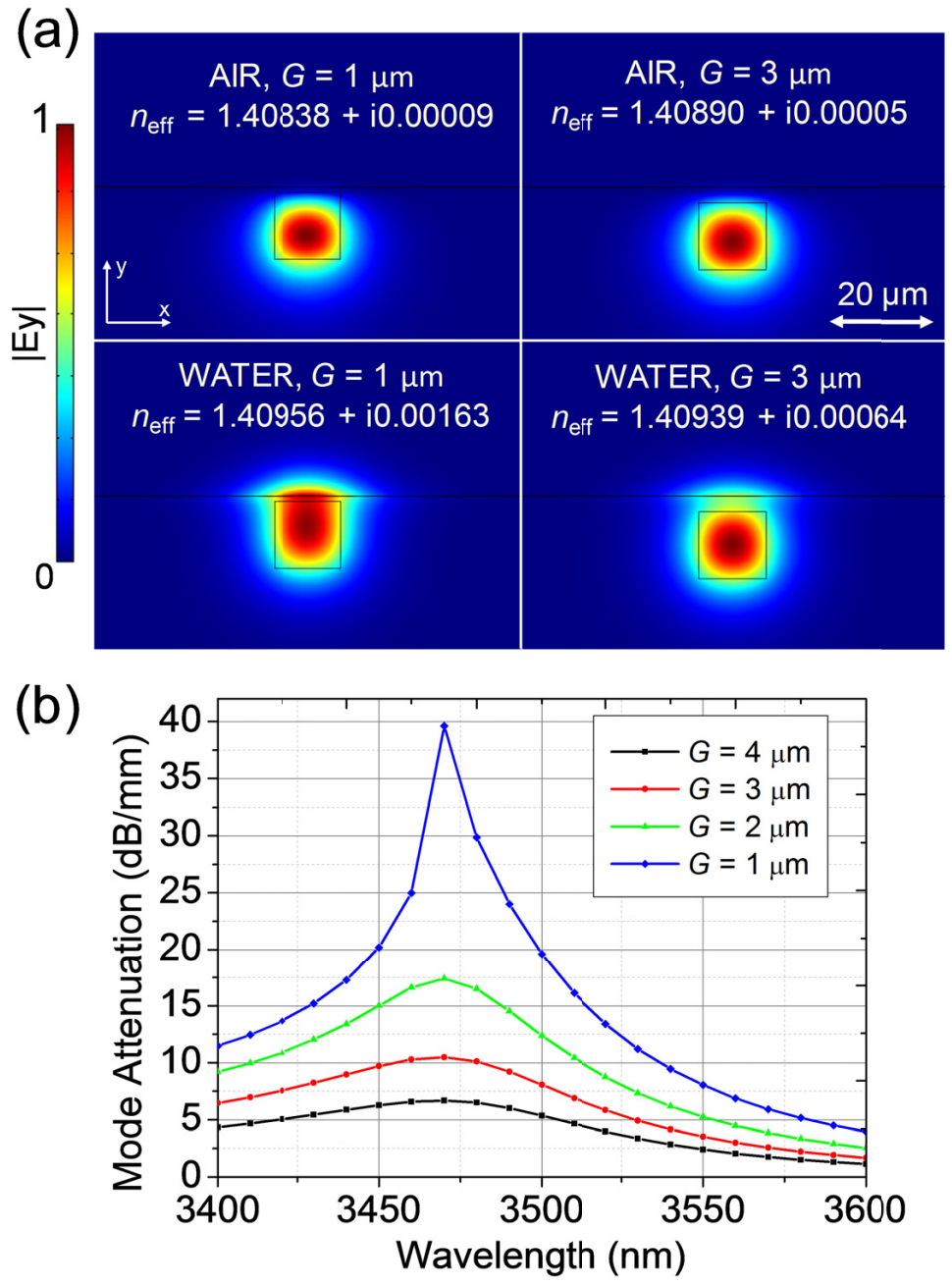

**Fig. 3.** (a) Simulated TM electric field profile magnitude at $\lambda = 3.49$ µm for different superstrates and gaps. (b) Calculated SPP relative mode attenuation (*MA*) spectrum for different gaps.

From the simulated effective extinction coefficients one can calculate the relative mode attenuation (*MA*) which is here defined as $MA\ (dB/mm) = 4\pi \cdot \log(e) \cdot 10^4 (k_{water} - k_{air})/\lambda(\mu m)$, where $k_{water}$ and $k_{air}$ are the mode extinction coefficients when the *superstrate* is water and air, respectively, and $\lambda$ is the free-space wavelength in mm. The *MA* spectrum for the SPP coupled mode is shown in Fig. 3(b) for various gaps *G*. As mentioned, the waveguide-SPP coupling and, therefore, the attenuation is stronger for shorter gaps leading to a decreasing width of the resonance which makes the device more sensitive and selective. The maximum *MA* is located at $\lambda = 3.47$ µm which may not necessarily coincide with the maximum measured attenuation.

For the correct experimental demonstration of the SPP excitation, a broadband mid-IR source is needed. A home-built high brightness fiber-based IR supercontinuum source (SCS) was therefore used [16]. The small core (7.5 µm) and NA (0.265) of the fiber guarantees low-loss mid-IR coupling with the waveguide. The resulting 2.4 to 4.1 µm spectrum (see Figure 4) of the light generated by the SCS was recorded by a Jobin-Yvon Horiba HR460 monochromator equipped with a 300 gr/mm IR ruled grating and having a PbSe detector at the output slit. The detection system was synchronized to the seed laser repetition rate (5.12 kHz) by means of a lock-in amplifier for maximum SNR.

Because only the longest wavelengths (> 2.4 µm) are relevant, a germanium-based long-pass filter was placed between a pair of aspheric lenses that collimated the light from the SCS and coupled the filtered light to another single-mode transporting fiber. This also protects the chip from the high intensity residual seed at $\lambda$=1.55 µm that had been greatly amplified in the SCS. The measured average output power of the transporting fiber is 500 µW. The transporting fiber and an output fiber were carefully aligned to their corresponding sensor chip ports. When the waveguide insertion loss was minimized, the second end of the output fiber was placed in front of the input slit of the monochromator in order to measure relative spectral variations of the light emerging from the mid-IR SPP sensor. Note that no polarizing elements were used here so no isolation of TM light was performed.

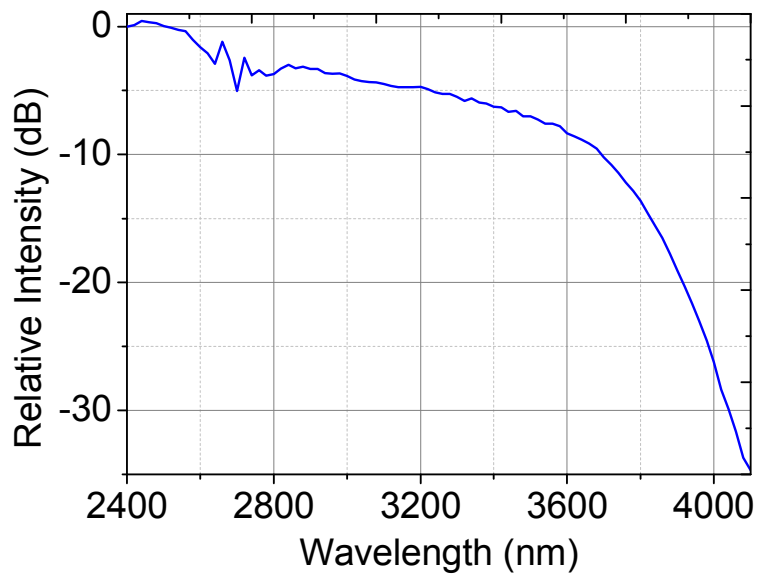

Fig. 4. Long-wavelength spectrum generated by the home-built SCS.

Spectrums with and without water drops covering the ITO thin film were measured so as to evaluate the relative light attenuation (*LA*) associated with a SPP excitation. Precisely, *LA* (dB/mm) = -10log($V_{water}$/$V_{air}$)/*L*, where $V_{water}$ is the detector voltage when a water drop is covering the sensor, $V_{air}$ is the clean reference voltage and *L* is the drop length (in mm) which acts as a normalization factor. *L* was determined by taking a top view image of the analyte drop, which introduced a ±50 µm uncertainty in the measurement (a microfluidic PDMS cover would eliminate this uncertainty but is also out of the scope of the present letter). An example of a *LA* spectrum using pure deionized water at 20 °C is shown in Figure 5(a). The curve has a resonant-like behavior which peaks at $\lambda$ = 3510 ± 10 nm, thus hinting a SPP excitation.

Direct analogy with the simulated *MA* (dashed line in Fig. 5(a)) may be done cautiously and taking into account that in the experimental setup the polarization is not controlled, and that the device transmission not only depends on the polarized modal attenuation but on other processes such as the mode coupling efficiency dispersion between the pure waveguide mode and the hybrid waveguide-SPP mode. These factors may explain the lower attenuation obtained in this SCS experiment, that is 0.9 dB/mm versus 9.8 dB/mm (simulated) TM mode attenuation. The parameters used for this simulation were $W \times H$ = 11.8 × 12.7 µm², $\Delta n = 8\cdot 10^{-3}$, $G$ = 3.3 µm and $TH$ = 54 nm, which were extracted from the previous characterization of the fabricated device.

Following the same procedure that with pure water, drops of a solution of water and ethanol at 20 °C were tested on the SPP sensor. In Fig. 5(b), the measured spectrums for different volume concentration of ethanol in water (from 0 % to 50 %) are shown. Differences in several CH and OH bands for increasing ethanol concentration are observed. For instance, a rise of LA occurs around $\lambda \approx 3.3$ µm and, more intensively, from $\lambda \approx 3.6$ to 3.8 µm. These spectral variations can be associated to the increasing presence of the CH stretching band around ~2970 cm$^{-1}$ [17], which can produce SPP coupling at different wavelengths due to absorption increments or index increments, i.e. both real and imaginary parts of the dielectric and the hybrid-SPP mode effective indices affect the coupling ratios and hence the measured attenuation.

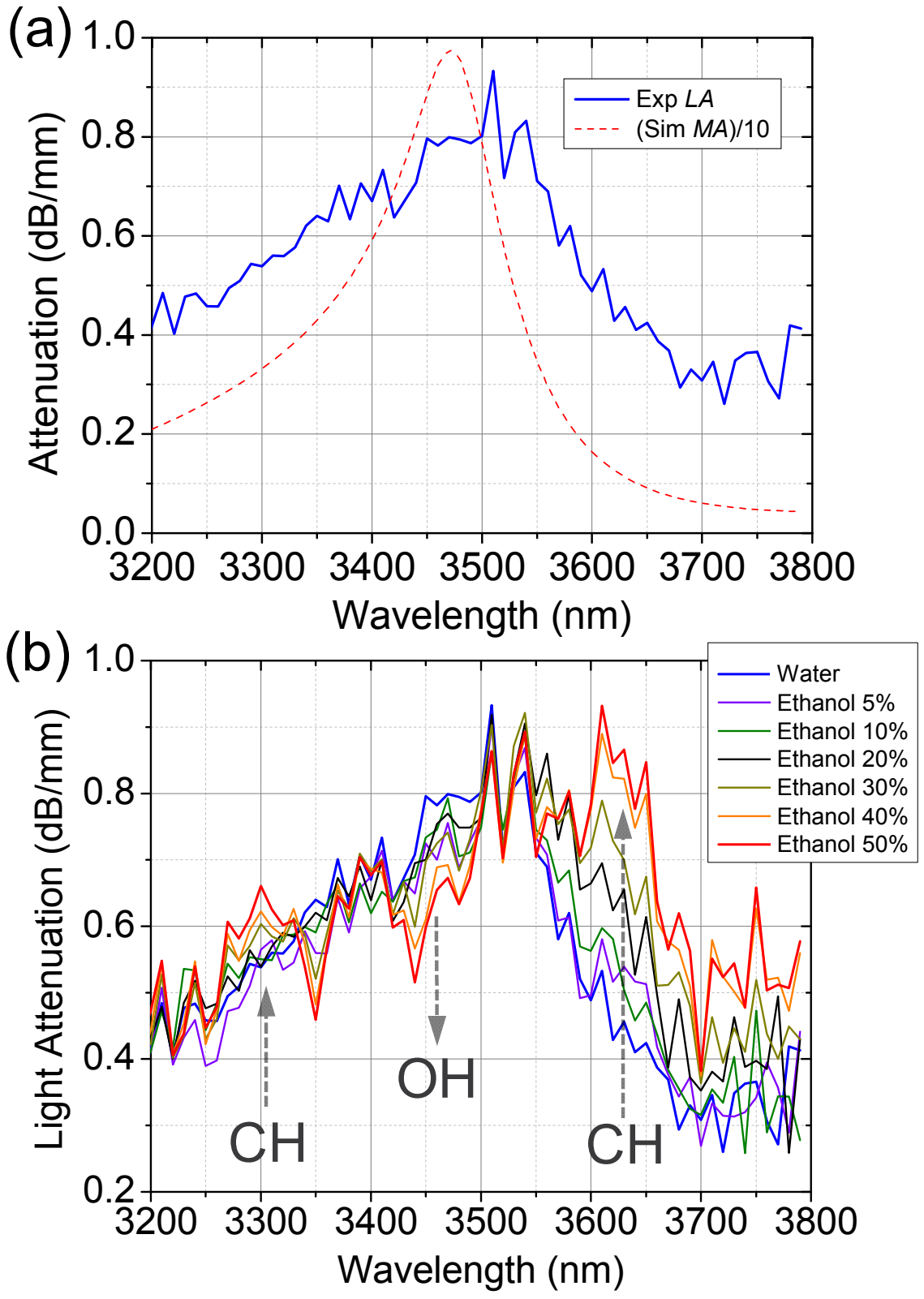

Fig. 5. (a) Unpolarized light attenuation (*LA*) spectrum (solid line) when a drop of water is placed on the sensor. The calculated TM *MA* spectrum (divided by 10) is superimposed (dashed line) for comparison purposes. (b) Unpolarized *LA* spectrums corresponding to different solutions of water and ethanol with increasing ethanol concentrations.

The higher attenuation at $\lambda$ ~3.3 µm is directly produced by a stronger optical absorption (CH stretch), whereas the growing attenuation in the 3.6 – 3.8 µm band is due to the refractive index increase (around the anomalous dispersion range) associated with the CH resonance. As already explained for the pure water case, such index increase facilitates the symmetry condition for a SPP excitation and the modes coupling, with the difference that for the CH stretch it happens at a longer wavelength as it can be clearly observed in the new peak emerging at $\lambda$ ~3610 nm, for the higher concentration case (30 – 50 %). Conversely, at the same time the *LA* at $\lambda$ ~3450 nm is reduced indicating a diminishing presence of the OH's against CH's within the SPP near-field. In short, the complex refractive index of the water-ethanol solution has a spectrally complex nature in this region which allows for SPP excitation around two distinct bands corresponding with the OH and CH stretching molecular vibrations.

The mid-IR SPP sensor was further tested for the water-ethanol specific case by using a 3.60 µm narrow-band low-noise interband cascade laser (ICL) in a TM-polarized setup, therefore allowing for a much more precise reading of the SPP effect. For these tests the waveguide output was imaged into a FLIR SC7000 mid-IR camera which allowed for direct monitoring of the output power temporal evolution. Polarizers in TM direction were placed before and after the waveguide to isolate the SPP effect. The same experiment as with the SCS was repeated but adding low ethanol concentration samples down to 1%.

As shown in Figure 6(a), an increasing *LA* associated to the CH feature at 3.60 μm for higher ethanol concentration was observed in the spectral measurements. The included ±2 % relative error bars corresponds to the ±50 μm uncertainty in the drop size determination. Furthermore, the maximum ethanol measurable concentration was estimated to be at least of 50 %, having a higher slope (i.e. higher sensitivity) in the range 0 - 30 %, and a decreasing one from 30 % to 50 %. As expected, higher attenuation is obtained than in the preceding experiment since only TM light was coupled into the waveguide. At 0% ethanol (i.e. pure deionized water) the experimentally measured *LA* is 1.75 dB/mm, in very good agreement with the simulated *MA* at $\lambda$ = 3.60 μm of 1.63 dB/mm. It was also verified that *LA* is negligible when launching TE polarized light regardless the analyte because no TE SPP mode can exist.

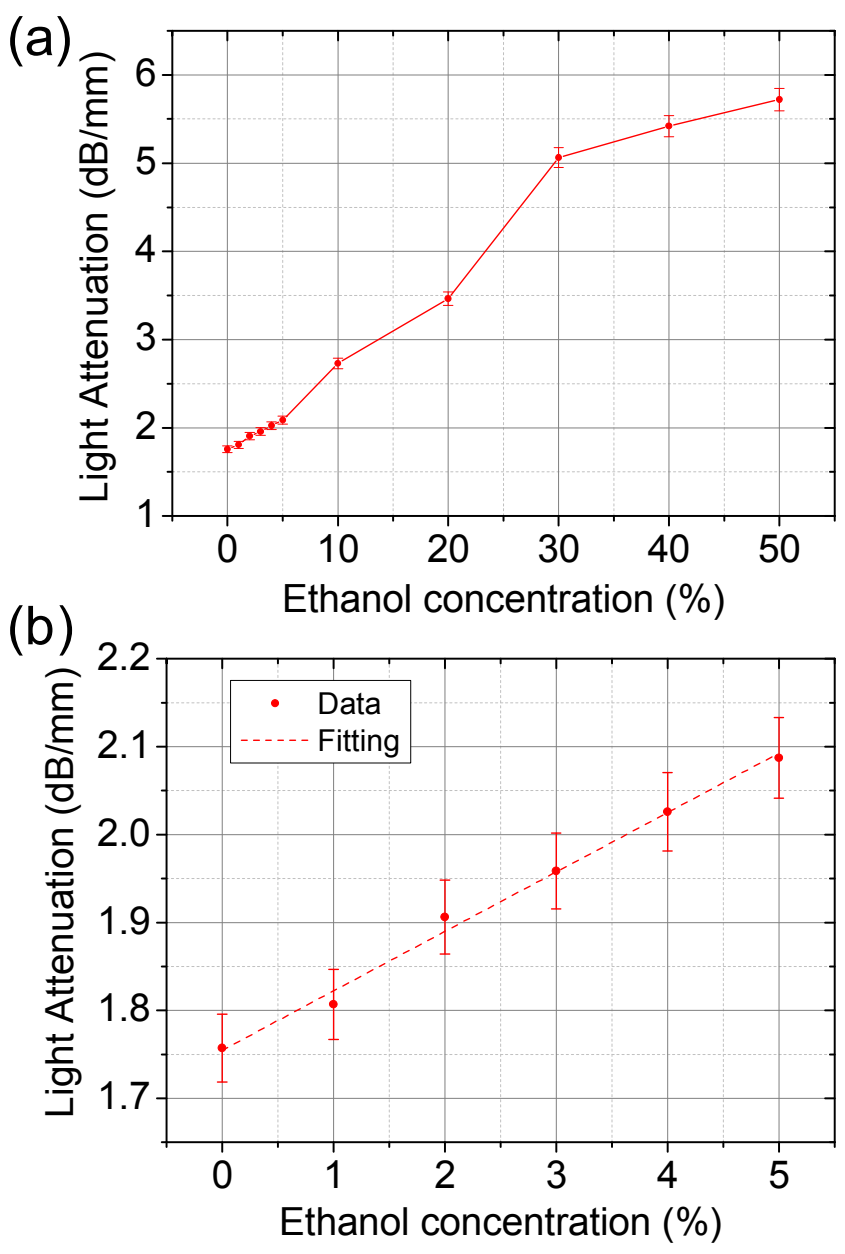

Fig. 6. (a) TM polarized *LA* induced by the water-ethanol solution drops on the sensor at $\lambda$ = 3.6 μm for increasing ethanol concentrations. (b) *LA* for low ethanol concentrations and linear fitting of the data.

Focusing on the lower concentrations, Fig. 6(b) shows the measured *LA* from which a clear linear trend can be deduced. Using this as a calibration curve, and using mean and standard deviation values derived from the several measurements, the instrumental limit of detection (*LOD*) of ethanol of this SPP sensor can be estimated [18]. It is important to note that for this calculation the *L* uncertainty (drop length) of our laboratory experiments is neglected, since adding a microfluidic channel on the ITO surface would allow to easily fix the interaction length. For *L* = 2 mm, 3$\sigma$ criterion, and 10 measurements, the *LOD* is calculated to be 0.02 %. The *LOD* of this mid-IR SPP sensor matches that of other sensors fabricated with more complex techniques like the ones found in [4] (0.06 %) and [19] (0.004 %), which are based on plasmonic and capacitive MEMS technology, respectively.

In conclusion, the excitation of a mid-IR SPP on an ITO thin film covered by an aqueous analyte via a 3DLW silica glass waveguide has been demonstrated. The device has been modeled trough FEM simulations, then fabricated with well-stablished microfabrication techniques and finally experimentally validated using broad and narrowband spatially coherent light. The attenuation due to the SPP excitation has been used to sense the presence of water and the concentrations of ethanol in water, taking advantage of the anomalous dispersion range of the liquids at their molecular resonances. The highly efficient mid-IR fiber-to-SPPs coupling on common ITO coated silica glass could find wide applications in biological and chemical sensing and could be directly implemented in portable smart phone screens, microscopy slides, or any other glass surface used in everyday life.

**Funding.** Spanish Government (MAT2013-47395-C4-4-R, TEC2014-55948-R, TEC2014-52642-C2-1-R); Catalan Government (2009SGR235, 2014FI_B00274); European Commission (EC) (ACP2-GA-2013-314335-JEDI ACE).

**Acknowledgment**. F. D. acknowledges additional support 2010-ICREA-02 for excellence in research.

## References for reviewing